\documentclass[a4paper]{raa}
\usepackage{graphicx,times}
\usepackage{natbib}
\usepackage{amssymb,amsmath}
\bibpunct{(}{)}{;}{a}{}{,}
\usepackage[dvipdfm,
			pagebackref=true]{hyperref}

\hypersetup{pdftitle={Uncertainties in measuring of scattering screens distances in pulsars observation}, 
			pdfauthor = {Evgeny Fadeev}, 
			pdfsubject= The subject, 
			pdfkeywords = keyword1 keyword2 keyword3} 
\hypersetup{colorlinks = true,
			linkcolor = green, 
			anchorcolor = red, 
			citecolor = blue, 
			filecolor = red, 
			urlcolor = red}

\begin{document}

\title{Uncertainty in Measurements of the Distances of Scattering Screens in Pulsar Observations}

\volnopage{ {\bf 2012} Vol.\ {\bf X} No. {\bf XX}, 000--000}
\setcounter{page}{1}

\author{Evgeny N. Fadeev\inst{1}}
\institute{ 
Lebedev Physical Institute, Astro Space Center, Profsoyuznaya 84/32, Moscow 117997, Russia; {\it fadeev@asc.rssi.ru}\\
\vs \no
{\small Received 2017 October 30; accepted 2017 October 30}
}

\abstract{In our previous paper we investigated properties of the ionized interstellar medium in the direction of three distant pulsars: B1641-45, B1749-28 and B1933+16. We found that uniformly distributed scattering material cannot explain measured temporal and angular broadening. We applied  a model for a thin scattering screen and found the distances to the scattering screens in all directions. In this paper, we consider more complicated models of scattering material distribution, such as models containing both a uniformly distributed medium and thin screen. Based on these models, we estimate the accuracy of localization of scattering screens and the possible relative contribution of each scattering component.
}

   \authorrunning{E. N. Fadeev et al. }            
   \titlerunning{Measurements of distances of scattering screens}  
   \maketitle

%
\section{Introduction}           
\label{sect:intro}
Inhomogeneities in the interstellar plasma distort radio emissions, which leads to angular broadening of radio sources. The angular diameters of scatter-broadened high-latitude ($|b| > 10^\circ$) extragalactic sources decrease with galactic latitude (\cite{1972Natur.236..440R}), whereas for low-latitude sources ($|b| < 10^\circ$) there is a tendency for them to be larger the closer they are to the Galaxy's center (\cite{1984Natur.312..707R}). In contrast, \cite{2008ApJ...672..115L} found that there is no correlation between the galactic latitude and the scattering diameter in the direction of the anticenter of the Galaxy that can be explained by a small quantity of uniformly distributed scattering material in the external part of the Galaxy.

\cite{Gwinn1993} showed that temporal and angular broadening of the radiation from pulsars corresponds to scattering by a uniform medium, except for several young pulsar are scattered by their supernova remnants. The same conclusion was made by \cite{Britton1998}. \cite{Stinebring2001} found  parabolic structures in the secondary spectra (a secondary spectrum is the power spectrum of the dynamic spectrum) of some pulsars that arise as a result of scattering by material situated in compact areas (known as thin screens) in the line of sight. 

The ground-space radio interferometer RadioAstron has a sufficient angular resolution to resolve scattered discs of some pulsars, which allows us to study the distribution of scattering matter in detail (\cite{Gwinn2016}, \cite{Popov2016}, \cite{Popov2017a}, \cite{Shishov2017}, 
\cite{Andrianov2017}).

\section{Models of Distribution of Scattering Matter}
\label{sect:tcmm}
\subsection{Simple Scattering Models}
In our previous paper (\cite{Popov2016}), we measured independently both the angular diameter of the average scattering disk $\theta_H$ and the temporal broadening time $\tau_{sc}$. Dependences of these values on mean scattering angle per unit length $\psi(z)$ have been given by expressions 
(\cite{1985MNRAS.213..591B})
\begin{equation}
\label{fo:1}
\theta_H^2 = \frac{4 \ln{2}}{D^2} \int\limits_0^D z^2 \psi(z)dz
\end{equation}
\begin{equation}
\label{fo:2}
\tau_{sc} = \frac{1}{2cD} \int\limits_0^D z \left ( D - z \right) \psi(z)dz
\end{equation}
where z is a coordinate along the line of sight from source ($z = 0$) to observer ($z = D$), and $c$ is the speed of light. \cite{Gwinn1993} (see also \cite{Britton1998}) have considered two important cases: uniformly distributed scattering material ($\psi(z) = \Psi_{0}$) and scattering material concentrated in the thin screen ($\psi(z) = \Psi_{1} \delta (z - (D - d_{s}))$). Here $d_{s}$ is the distance from the observer to the screen, and $\Psi_0$ and $\Psi_1$ are constants. Following this approach, we have shown that measured $\theta_H$ and $\tau_{sc}$ do not satisfy the relationship for a uniform medium which is $\theta_u^2 = 16 \ln 2 c \tau_{sc} / D$. Using the thin screen model, we have calculated screen distances for three pulsars. 

It is clear that the interstellar medium has a more complex composition than a thin single layer on the way to a pulsar. Structures that scatter radio waves might be localized mainly into areas with high turbulence, such as shells of H\,II regions, whose sizes usually are only a small part of the distance to the pulsars. Several ionized clouds with a different scattering power may be located along the line of sight. Uniformly distributed interstellar medium in the long distances also may produce significant scattering. The same temporal and angular broadening can be produced not only by a single screen but also by several thin screens with different scattering power or by a mixture one or more screens, immersed in the uniform medium. In this regard, it seems important to consider more sophisticated models of the interstelar medium to figure out if there is a dominant scattering screen. If such a screen exists, we need to assess the accuracy of localization for this screen under the assumption that other scattering agents exist.

\subsection{Thin Screen and Uniform Medium}
\cite{Gwinn1993} proposed a two-component model that includes scattering by both uniformly distributed material and a thin screen. For such a model, 
$$\psi(z) = \Psi_1 \delta(z - (D - d_s)) + \Psi_0.$$
In this case, the relationship between $\theta_H$ and $\tau_{sc}$ is as follows:
\begin{equation}
\label{fo:3}
\theta_{H}^2 = \theta_{u}^2 \frac {\chi + 3s^2}  {\chi + 6s \left (1-s \right )},
\end{equation}
where $\chi = \Psi_0 D / \Psi_1$ is the relative power of scattering and $s = ( D - d_s ) /D$ is the fractional distance of the screen from the pulsar to the observer. When all scatterings are produced by the screen ($\Psi_0 D \ll \Psi_1$ and $\chi \to 0$), we obtain the expression describing the thin screen alone, $\theta_H^2 = \theta_u^2 s / (1-s) / 2$. For $\chi \to \infty$ we return to a uniform scattering material distribution. It is important that if the scattering screen is located at the distance $s = 2/3$, it produces the same size of scattering disk as the uniformly distributed medium, and in the absence of additional data we cannot distinguish these two cases.

\begin{figure}[h!]
   \centering
   \includegraphics[width=14.0cm, angle=0]{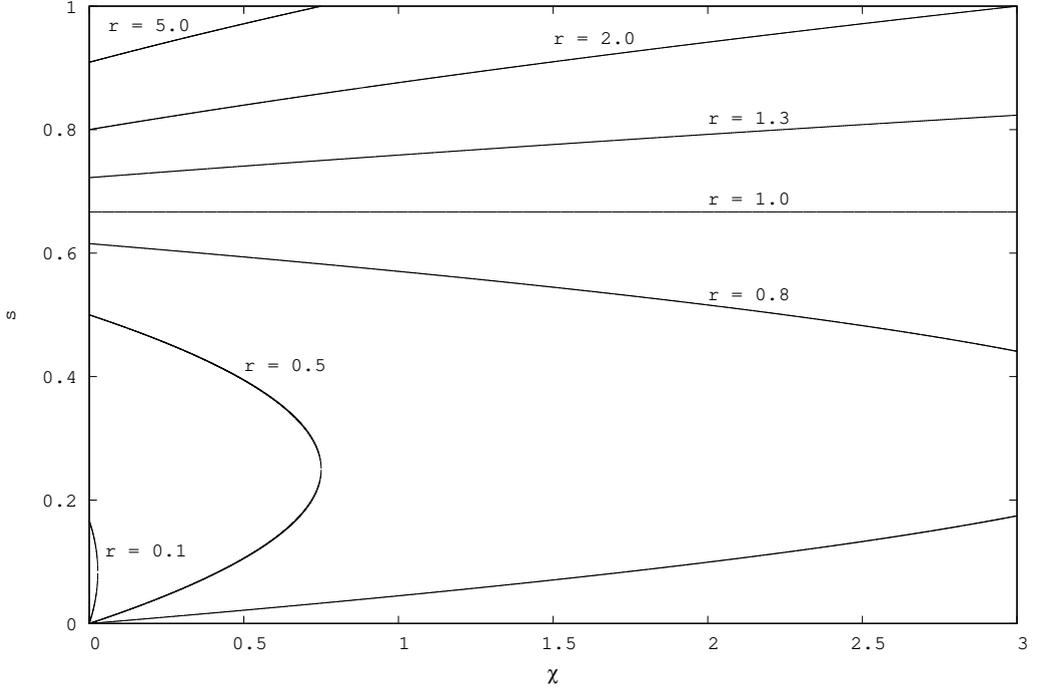}
   \caption{The dependence of fractional distance $s$ of the scattering screen plotted with the contribution of uniform medium $\chi$ for different measured relative sizes of scattering disk $r$. There are two groups of solutions. If the screen position in the uniform medium free model $s(\chi = 0) = s_0$ is less than $2/3$, the real position can be at any distance from pulsar to $s_0$. The maximum contribution of the uniform scattering medium occurs for the distance $s_0 / 2$. For screens with $s_0$ more than $2/3$, the real screen position can be anywhere between $s_0$ and the observer.} 
   \label{Fig:1}
\end{figure}

We can solve  equation [\ref{fo:3}] for $s$
\begin{equation}
\label{fo:4}
s = \frac{1 \pm \left[1-\frac{(1+2r)(1 - r)}{3r^2}\chi\right]^{1/2}} { 2 + r^{-1} },
\end{equation}
where $r = \theta_{H}^2 / \theta_{u}^2 $ is the relative size of the scattering disk. Figure [\ref{Fig:1}] shows the fractional distance $s$ plotted against the relative power of scattering $\chi$. For $r \in \left [0;1 \right ) $ there are solutions with $\chi \in \left [0;  \chi_{max} \right ]$. Here $\chi_{max} = {3r^2} \left[(1+2r)(1 - r)\right]^{-1}$ is the greatest possible contribution of the uniform medium along the line of sight in the total scattering for measured values of  $\theta_{H}$ and $\tau_{sc}$. If the influence of the uniform medium is negligible ($\chi = 0$), the position of the scattering screen is the farthest from the pulsar and corresponds to the single screen case: $s_0 = (1+(2r)^{-1})^{-1}$. But if the uniform medium also contributes to the scattering, the screen stands closer to the pulsar. When the extended medium scatters the radio emission with maximum strength ($\chi = \chi_{max}$), the distance to the thin screen is decreased down to $s_m = s_0 / 2$. Significantly, (\ref{fo:4}) provides two possible values for $s$, even when the extended medium is absent. One of them is $s_0$ and the other one is $0$. If $\chi$ is not 0, the first value of $s$ is less than $s_0$ and it decreases with increasing $\chi$. At the same time, the second possible value of $s$ is increasing with $\chi$ up to $s_m$.  

So if we take into account the uniform medium, we can place the scattering screen anywhere between its location corresponding to the case without a uniform medium and the pulsar. Moreover, we cannot clearly distinguish two possible screen positions even if the value of $\chi$ is known.

For $r > 1$ there is the only possible value of $s$ that equals $s_0$ in the single screen model, which increases to 1 with the scattering strength of the uniform medium ($\chi$). Since for $r > 1$ and $s > 2/3$ if we find out the screen distance $s_0$ is more than 2/3, the presence of additional uniformly distributed material shifts the position of the scattering screen closer to the observer.

Thus we can locate the position of the scattering screen with certainty when it is close to the pulsar or the observer. In other cases, the true position of the screen can noticeably differ from the derived one for the single screen model.

\subsection{Two Scattering Screens}
\citet{Putney2006} have shown that the secondary spectra of some pulsars exhibit multiple scintillation arcs with different curvatures. In our paper \citep{Popov2016}, we have reported the detection of two parabolic arcs in the spectrum of pulsar B1933+16. Each single arc is produced by a separate thin screen whose location can be determined if the pulsar distance and its proper motion are known.  For two scattering screens and uniform medium,
$$\psi(z) = \Psi_0 + \Psi_1 \delta(z - (D - d_1)) + \Psi_2 \delta(z - (D - d_2)),$$
and ratios  $\chi_1 = \Psi_0 D / \Psi_1$ and $\chi_2 = \Psi_0 D / \Psi_2$ of scattering strength for uniform medium and screens are bounded by the following relation
\begin{equation}
\label{fo:5}
\chi_1 = \frac{3s_1 \left [2 r (1 - s_1) - s_1 \right]} { 1-r - 3 s_2 \left [2r(1-s_2) - s_2  \right] /\chi_2}.
\end{equation}
If determination of screen positions by means of scintillation arcs is accompanied by measurements of angular and temporal broadening of pulsars, we can estimate the relative scattering power of thin screens and the uniform medium. Neglecting the uniform medium, we obtain ratio $\chi_{12} = \Psi_1 / \Psi_2$ as the function of $r$, $s_1$, and $s_2$
\begin{equation}
\label{fo:6}
\chi_{12} = \frac{s_2}{s_1} \frac{2( 1-s_2 )r - s_2} { s_1 - 2(1-s_1)r }.
\end{equation}

It must be emphasized that observable angular and temporal broadening is produced by all screens and the uniform medium. The measurement of only $\theta_{H}$ and $\tau_{sc}$ does not make it possible to choose the single-component or the multi-component interstellar medium model.

\section{Comparison with observations}
\label{sect:comparison}
We estimated the position of thin scattering screens in the directions of three pulsars: B1641-45, B1749-28 and B1933+16 (\cite{Popov2016}). For B1641-45, our estimates of $s$ were $0.36 \pm 0.02$ for a distance to the pulsar of $4.5 \pm 0.4$\,kpc. According to [\ref{fo:4}], we can infer that the scattering screen is located somewhere between $s = 0$ and $s = 0.36$, i.e. the screen distance varies from 2.7 to 4.9\,kpc depending on the pulsar distance. The model allows the maximum value of $\chi_{max} = 0.20 \pm 0.03$, i.e. a thin screen scatters radio waves at least five times more strongly than a uniform medium. The H\,II region G339.1--0.4 lies close to the line of sight at distance 3.3\,kpc. If we assume that this H\,II region acts as the scattering screen, we can estimate the contribution of uniform medium $\chi$ as 0.14.    

The distance to the pulsar B1749-28 is  not well-known: $0.2_{-0.1}^{+1.1}$\,kpc (\cite{Verbiest2012}). At the lower limit of this distance, the contribution of the uniform medium is negligible, and the scattering screen is located very close to the pulsar. For the upper limit the maximum contribution of a uniform medium still remains small but the screen can be shifted up to 300\,pc from the pulsar.      

In the secondary spectrum of B1933+16, there is a complex pattern that includes at least two parabolic arcs. Each arc corresponds to one scattering screen. We have located these screens at distances $1.0-1.1$\,kpc and $2.4-3.7$\,kpc from the observer for pulsar distance $3.7_{-0.8}^{+1.3}$ (\cite{Verbiest2012}). The analysis of temporal and angular broadening gives the screen position $2.3 - 3.4$\,kpc in the single screen approach. Neglecting the uniform medium described by [\ref{fo:6}], the more distant screen scatters radio waves from $20$ to $40$ times more strongly than the closer one. This result  was expected because the position of the more distant and powerful screen is in good agreement with the one scattering screen model. Equation [\ref{fo:5}] allows the existence of a uniform medium that scatters 10-30 times ($\chi_2 = 0.03 - 0.09$) less strongly than the  more powerful screen. At higher values of $\chi_2$, the weaker screen
cannot exist.

\section{Conclusion}
\label{sect:conclusion}
We have shown that uniformly distributed scattering material affects the location of the thin scattering screen. Without additional data, the real position of the scattering screen remains unknown and lies between the position  calculated without the uniform medium and the pulsar ($d_s > D/3$) or the observer ($d_s < D/3$).

The uniform medium is responsible for no more than $20\%$ of all scattering in the directions of pulsars B1641-45, B1749-28 and B1933+16. In the direction of the pulsar B1933+16, material that causes scattering is concentrated in two thin screens, with one of them being at least 10 times more powerful than the other.

\normalem
\begin{acknowledgements}
I thank M.~V.~Popov,  A.~S.~Andrianov, A.~G.~Rudnitskiy, T.~V.~Smirnova and V.~A.~Soglasnov for useful discussions.

RadioAstron is a project of the Astro Space Center of the P.~N.~Lebedev Institute of the Russian Academy of Sciences and the S.~A.~Lavochkin Research Industrial Association, under contract with the Russian Space Agency, together with many science and technological organizations in Russia and other countries.
\end{acknowledgements}

\bibliographystyle{raa}
\bibliography{biblio}

\end{document}